# Khristianovich-Geertsma-de Klerk problem with stress contrast

I. O. Gladkov[1], A. M. Linkov[1,2]

*Institute for Problems of Mechanical Engineering, 61, Bol'shoy pr. V. O., Saint Petersburg, 199178, Russia*
*Rzeszow University of Technology, ul. Powstancow Warszawy 8, Rzeszow, 35-959, Poland*
*e-mail: linkoval@prz.edu.pl*

*Abstract.* The paper contains an extension of the Khristianovich-Geertsma-de Klerk (KGD) model to the case when the confining rock pressure, which closes a hydraulic fracture, varies in the direction of its propagation. The extension is impelled by the need to simulate fracture hampering (acceleration) when it penetrates into a layer with increased (decreased) rock pressure. The paper presents the problem formulation, an efficient numerical method for its solving, examples of fractures propagating through layers with various stresses and general conclusions. It is established that when the fracture enters a layer with increased rock stresses (positive stress contrast), it actually stops. In this case, the fluid particle velocity drops practically to zero and the velocity near the tip oscillates about a small value until the fluid pressure increases to the level of the increased rock pressure. In the opposite case, when the fracture enters a layer with decreased stresses (negative stress contrast), the fracture front accelerates until the fluid pressure drops to the level of the decreased rock pressure. Physical considerations and numerical results imply that the transition through a boundary of layers with different stresses may be characterized by a simple dimensionless parameter. The latter is defined as the ratio of the average difference between the fluid and rock pressures (at the moment of reaching the contact) to the stress jump at the contact. The parameter is applicable to contacts with positive as well as negative contrasts. The model and the method developed are to serve for improvement of the popular pseudo three-dimensional (P3D) model.

## 1. Introduction

Hydraulic fracturing (HF) is an operation widely used for stimulation of fluid flow to a borehole. It is also applied in geothermal technologies, sequestration of toxic waste and $CO_2$, roof control and preventing rockbursts in mines (see, e. g. reviews in papers [1,2]). In view of limited and uncertain information on the properties and the state of rock in a pay layer and embedding mass, the analytical study and numerical modeling are of significance for success of this expensive operation [3]. Mathematical investigations were started by Khristianovich and Zheltov [4,5]. The authors assumed that a cross section orthogonal to the fracture front was in the plane-strain state. Later Geertsma and de Klerk developed the model in the form known presently as the Khristianovich-Geertsma-de Klerk (KGD) model. Since then, investigations of HF have got growing attention (e.g. [7-28]). Despite of huge number of publications on simulation of HF, the KGD model is still of principal significance due to its pioneering nature and to the facility to obtain very accurate benchmark solutions (e.g. [16, 22,25]). Its importance has grown last years in view of yet existing difficulties in modeling 3D problems in real time what increased interest in the development of the pseudo three-dimensional (P3D) model. The latter is based on the Perkins-Kern-Nordgren (PKN) model [7,9], in which plane-strain conditions are assumed in cross-sections parallel to the fracture front, the fluid pressure is assigned constant in a cross section and rock mass is considered elastic with the homogeneous confining pressure $\sigma$ closing a crack. The extension of the PKN model to the P3D model [10,3,20,24] includes two items. Firstly, the confining stress $\sigma$ is assigned as a step function along the fracture height what serves to account for different rock pressure in different layers. Secondly, in contrast with the classical PKN model, the fracture height is not constant; rather it grows with some speed. Assigning this speed is a tricky issue, which is tackled by assuming various simplifications. For instance, in the paper [3], its authors assume that the stress intensity factor at the fracture tips equals to a critical value $K_{IC}$. However, in many problems, for instance in the case, when there is no stress contrast, this yields to inexistence of a solution for the P3D model. To avoid this flaw, a fictitious coefficient $K_{IC}$ is introduced [3,10,24], which is chosen to make results acceptable. In papers [3,10], a notice was made on attempts to assign the fictitious $K_{IC}$ by using the KGD model. Meanwhile, except for the paper [24], no clear explanation how to do it has been given.

The authors of the paper [24] considered a problem for a pay layer located between half planes with the same stress contrast. The hydraulic fracture propagates symmetrically about the middle of the pay layer, thus the problem is symmetric with respect to the fracture center. For this particular case, the KGD problem is solved with stress contrast. Its solution is used for semi-empirical choice of $K_{IC}$ as a function of the fracture speed. The success reached for the particular case indicates that there is sense to solve the KGD problem for an arbitrary composition of layers with arbitrary stress contrasts. The objective of this work is to formulate, numerically solve and discuss general features of the generalized KGD problem.

The objective is reached on the basis of the modified theory of hydraulic fracturing summarized in [26]. We consider the general case, when a problem may be non-symmetric about the fracture center and, consequently, speeds of tips may differ. The problem formulation and the method developed account for the propagation of both tips. The method is applicable to problems with arbitrary stress contrast, which may be positive or negative and very high in absolute value. For the particular case considered in the paper [24], there is agreement of the results to the accuracy of graphs presented in [24]. From the numerical results it follows that there is the principal difference between the case when the fracture penetrates a layer with higher stresses (positive contrast) and the case when it enters a layer with lower stresses (negative contrast). In the first case, the fluid particle velocity drops practically to zero and the velocity near the tip oscillates about a small value until the fluid pressure increases to the level of the increased rock pressure. In the second case, when the fracture enters a layer with decreased stresses (negative stress contrast), the fracture front accelerates until the fluid pressure drops to the level of the decreased rock pressure. We also establish a dimensionless parameter, which characterizes peculiarities of the transition through a boundary of layers with positive, as well as negative, stress contrast.

## 2. Problem formulation

Straight hydraulic fracture of the length $2x_*(t)$ grows under conditions of plane strain in the direction orthogonal to the direction of compressive stresses $\sigma(x)$ in a layered mass containing $K$ layers (Fig. 1). The layers are numerated upward so that the bottom layer has the number 1, while the top $K$. The region is chosen thick enough to be sure that the areas below the first and above the last layer are certainly beyond the threshold of possible crack growth. Then the first and the last layers may be represented by half-planes. Layers between them have finite thickness $h^j$ ($j = 2,…,K$-1). The upper boundary of the $j$-th layer is assigned with the number $j$, thus the contact between layers $j$ and $j$ + 1 has the number of the lower layer $j$ ($j = 1,…,K$). The global $x$-axis is directed upward, and the origin is located at the center of the pay layer, which has the number $j_p$ and the thickness $h^{j_p} = H$. Therefore, the first boundary has the coordinate $x^1 = -\frac{H}{2} - \sum_{s=2}^{s=j_p-1} h^s$, while the following ones have coordinates $x^j = x^1 + \sum_{s=2}^{s=j} h^s$ ($j = 2,…,K$-1). The source of a fracturing fluid of the intensity

Fig. 1

$Q$ is located at the center of the pay layer ($x_Q = 0$). Similar to P3D model, the layers are assumed to be homogeneous with the same elasticity modulus $E$ and the Poisson ratio $v$ (see, e.g. [3]). The layered structure is taken into account via merely assigning various confining stresses $\sigma(x)$ acting along layers and closing the fracture. The presence of the contrast differs the problem formulation from the classical KGD model, in which the stress $\sigma(x)$ is constant along the direction of the fracture propagation. In the P3D model, the change of stresses in layers is prescribed as step-function with constant value of compression $\sigma^j$ at each of the layers:

$$\sigma(x) = \sigma^j \quad x^{j-1} < x < x^j, \quad j = 1, \dots, K \tag{1}$$

with $x^0 = -\infty$ and $x^K = \infty$. Note however that the derivative of step-function is the sum of delta-functions what in the considered modified KGD problem leads to computational difficulties. For this reason, below we consider the step-wise distribution of the closing pressure as limiting for continuous distribution $\sigma(x)$ and make a special study on this issue.

To simplify notation, the confining stress $\sigma(x)$, as well as the fluid pressure $p(x,t)$, are assumed positive. For linearly elastic rock, the fracture opening $w(x,t)$ is defined entirely by the difference $p_{net}(x,t) = p(x,t) - \sigma(x)$. It is convenient to write the latter as

$$p_{net}(x,t) = p_d(x,t) - \Delta\sigma(x) \tag{2}$$

where

$$p_d(x,t) = p(x,t) - \sigma_0, \quad \Delta\sigma(x) = \sigma(x) - \sigma_0 \tag{3}$$

Thus, in (2) $p_d(x,t)$ is the differential pressure, defined with respect to the stress $\sigma_0 = \sigma(0)$ in the pay layer, while $\Delta\sigma(x)$ is the "stress contrast" of the compressive stresses with respect to $\sigma_0$.

Rheology of the fracturing fluid is given by the power law (e.g. [3]) $\tau = M\dot{\gamma}^n$, where $\tau$ and $\dot{\gamma}$ are, respectively, the shear stress and the shear strain rate; $M$ is the consistency index, $n$ is the behavior index. For a Newtonian fluid, $M = \mu$ is dynamic viscosity, $n = 0$; for commonly used thinning fluids $0 < n < 1$.

By employing the modified HF theory [26], the mathematical problem is formulated as follows. For a non-compressive fluid, equations of volume conservation and momentum are written in terms of the particle velocity $v$:

$$\frac{\partial w}{\partial t} = -\frac{\partial(wv)}{\partial x} - q_l + \delta(x)Q(t) \tag{4}$$

$$v = -sign\left(\frac{\partial p_d}{\partial x}\right)\left(\frac{w^{n+1}}{\mu'}\left|\frac{\partial p_d}{\partial x}\right|\right)^{1/n} \tag{5}$$

where $\delta(x)$ is the Dirac's delta, $\mu' = \theta M$, $\theta = 2\left[\frac{2(2n+1)}{n}\right]^{1/n}$. The possibility to use in (5) $p_d$ instead of $p$ follows from equation (3), because the confining stress $\sigma_0$ in the pay layer is constant and, consequently, $\partial p/\partial x = \partial p_d/\partial x$.

The propagation of tips is defined by their speed equations:

$$\frac{dx_{*l}}{dt} = v_{*l} = \lim_{x \to x_{*l}} v(x,t), \quad \frac{dx_{*u}}{dt} = v_{*u} = \lim_{x \to x_{*u}} v(x,t) \tag{6}$$

where $x_*$ is the coordinate of a tip, $v_* = dx_*/dt$ is its speed; henceforth, the indices $l$ and $u$ denote respectively the lower and upper tip of the fracture.

The opening $w(x,t)$ is unknown and it depends on rock elasticity and the difference $p_{net}(x,t)$ between the fluid pressure and the confining stress. In view of (2), the hypersingular elasticity equation (e. g. [27]) is written as:

$$p_d(x,t) - \Delta\sigma(x) = \frac{E'}{4\pi} \int_{x_{*l}}^{x_{*u}} \frac{w(\xi,t)}{(\xi-x)^2} d\xi \qquad x_{*l} \leq x \leq x_{*u} \qquad (7)$$

where $E' = E/(1-\nu^2)$. A solution to (7) is searched in the class of functions $w$ equal to zero at tips:

$$w(x_{*l},t) = 0, \qquad w(x_{*u},t) = 0 \qquad (8)$$

The fracture propagation up and/or down is possible only when the stress intensity factor $K_I$ at the upper and/or lower tip reaches the critical value $K_{IC}$:

$$K_{Il} = K_{IC}, \qquad K_{Iu} = K_{IC} \qquad (9)$$

Henceforth, we consider the practically significant case (see, e.g. [17]), when the main resistance to the fracture propagation is defined by viscous shear at fracture surfaces. In this case the fracture toughness is taken zero: $K_{IC} = 0$. Then the opening and the differential pressure near fracture tips are defined by asymptotic equations for the viscosity dominated regime (e. g. [26]):

$$w(x) = A_w(v_*)r^\alpha, \qquad p_d(x) = -A_w(v_*)B(\alpha)E'r^{\alpha-1} \qquad (10)$$

where $A_w(v_*) = A_\mu(\alpha)(t_n|v_*|)^{1-\alpha}$, $A_\mu(\alpha) = [(1-\alpha)B(\alpha)]^{-\alpha/2}$, $B(\alpha) = \frac{\alpha}{4}\cot[\pi(1-\alpha)]$, $\alpha = 2/(n+2)$, $t_n = \left(\frac{\mu'}{E'}\right)^{1/n}$, $r = |x_* - x|$.

Employing (10) in (5) and (6) yields speed equations for tips expressed via asymptotics of the opening:

$$\frac{dx_{*l}}{dt} = v_{*l} = -\frac{1}{t_n}\lim_{x \to x_{*l}}\left(\frac{1}{A_\mu}\frac{w(x)}{r^\alpha}\right)^{\frac{1}{1-\alpha}}, \quad \frac{dx_{*u}}{dt} = v_{*u} = \frac{1}{t_n}\lim_{x \to x_{*u}}\left(\frac{1}{A_\mu}\frac{w(x)}{r^\alpha}\right)^{\frac{1}{1-\alpha}} \qquad (11)$$

As a result, we arrive at the complete system of five equations (4), (5), (7), (11) with five unknowns $w(x,t)$, $p_d(x,t)$, $v(x,t)$, $x_{*l}(t)$, $x_{*u}(t)$. The boundary conditions (8) and fracture conditions (9) under $K_{IC} = 0$ are met identically due to the asymptotic umbrella (10). The system contains temporal derivatives of $w$, $x_{*l}$ and $x_{*u}$. Thus initial conditions are needed for these quantities. They are:

$$x_{*l}(t_0) = x_{*l0}, \quad x_{*u}(t_0) = x_{*u0}, \quad w(x,t_0) = w_0(x) \quad x_{*l0} \leq x \leq x_{*u0} \qquad (12)$$

The problem consists of solving the system (4), (5), (7), (11) under the initial conditions (12) and with accounting for the asymptotic behaviour of the opening (10).

## 3. Normalized variables

It is convenient to exclude the elasticity modulus $E'$ and the viscosity parameter $\mu'$ from the equations by introducing the normalized opening [27]:

$$w_n = \left(t_s \frac{E'}{\mu'}\right)^{1/(n+2)} \qquad (13)$$

where the factor $t_s$ serves to conveniently choose an appropriate time scale. Then by using the normalized variables

$$w' = \frac{w}{w_n}, \; w'_0 = \frac{w_0}{w_n}, \; p' = \frac{p}{w_n E'}, \; \sigma' = \frac{\sigma}{w_n E'}, \; \sigma'^j = \frac{\sigma^j}{w_n E'}, \; p_d' = \frac{p_d}{w_n E'}, \; \Delta\sigma' = \frac{\Delta\sigma}{w_n E'},$$

$$q'_l = \frac{q_l}{w_n}, \; Q' = \frac{Q}{w_n} \tag{14}$$

we have the same equations for primed quantities while now $E' = 1$ and $\mu' = 1$. Emphasize that such normalizing does not influence the spatial coordinates and velocities. This serves to normalize the spatial coordinate in a way, which removes singularity of $\frac{\partial w}{\partial t}$ at the crack tips if the latter is evaluated under fixed value of the global coordinate. Since the tip speeds may differ both in the direction and magnitude, we introduce different normalizing for points below and above the source:

$$\varsigma = \frac{x}{|x_*|} = \begin{cases} \varsigma_1 = -x/x_{*l} & x \leq 0 \\ \varsigma_2 = x/x_{*u} & x \geq 0 \end{cases} \tag{15}$$

When $x$ runs from $x_{*l}$ to $x_{*u}$, the coordinate $\varsigma$ changes from -1 to +1. The temporal derivative $(\partial w/\partial t)_x$, evaluated under constant $x$, is connected with the temporal derivative $(\partial w/\partial t)_\varsigma$, evaluated under constant $\varsigma$, as $(\partial w/\partial t)_x = (\partial w/\partial t)_\varsigma + (d\varsigma/dt)\partial w/\partial \varsigma$. For the variable $\varsigma$, defined by (15), we obtain:

$$\left(\frac{\partial w}{\partial t}\right)_x = \left(\frac{\partial w}{\partial t}\right)_\varsigma - \varsigma \frac{v_*}{x_*} \frac{\partial w}{\partial \varsigma} \tag{16}$$

Since the opening is zero at the crack tips, the corresponding derivatives $(\partial w/\partial t)_\varsigma$ at them become zero, while the derivatives $(\partial w/\partial t)_x$ under the asymptotics (10) go to infinity in magnitude. This is an essential advantage of the variables $\varsigma_1$ and $\varsigma_2$.

In terms of the variables (14) and (15), after using (16) in the continuity equation (4), we obtain the system:

$$\frac{\partial w'}{\partial t} = \frac{1}{|x_*|} \frac{\partial [w'(\varsigma|v_*|-v)]}{\partial \varsigma} - \frac{v_*}{x_*} w' - q'_l + \delta(x)Q'(t) \tag{17}$$

$$v = -sign\left(\frac{\partial p'_d}{\partial \varsigma}\right) \left(\frac{w'^{n+1}}{\mu'} \left|\frac{1}{x_*} \frac{\partial p_d'}{\partial \varsigma}\right|\right)^{1/n} \tag{18}$$

$$\frac{dx_{*l}}{dt} = -\left(\frac{1}{t_s}\right)^{1/n} \lim_{\varsigma_1 \to -1} \left(\frac{1}{A_\mu} \frac{w(\varsigma)}{|x_{*l}(1+\varsigma_1)|^\alpha}\right)^{\frac{1}{1-\alpha}}, \; \frac{dx_{*u}}{dt} = \left(\frac{1}{t_s}\right)^{1/n} \lim_{\varsigma_2 \to 1} \left(\frac{1}{A_\mu} \frac{w(x)}{|x_{*u}(1-\varsigma_2)|^\alpha}\right)^{\frac{1}{1-\alpha}} \tag{19}$$

$$p_d'(\varsigma, t) - \Delta\sigma'(\varsigma) = \frac{1}{4\pi} \int_{-1}^{1} \frac{w'(\tau,t)}{|x_*(\tau)|[\tau - \varsigma |x_*(\varsigma)/x_*(\tau)|]^2} d\tau \qquad -1 < \varsigma < 1 \tag{20}$$

where it is assumed that $x_*(\eta) = x_{*l}$ when $\eta < 0$ and $x_*(\eta) = x_{*u}$ when $\eta > 0$. Pay attention that the coordinates of the tips, their speed and the particle velocities enter the system (17)-(20) non-normalized. They are measured in units conveniently chosen when introducing the normalizing (13), say in meters and meters per minute.

After normalizing the second of equations (3), the normalized contrast of stresses $\Delta\sigma'(\varsigma)$ for the step function (3) is:

$$\Delta\sigma'(\varsigma) = \Delta\sigma'^j = \sigma'^j(\varsigma) - \sigma'_0 \qquad \varsigma^{j-1} < \varsigma < \varsigma^j, \qquad j = 1, \ldots, K \tag{21}$$

where the normalized coordinates of layer boundaries, defining the stress contrast, depend on time:

$$\varsigma^j = \frac{x^j}{|x_*(x^j)|} \tag{22}$$

Near the tips ($\varsigma = \pm 1$), the asymptotic equation for the opening becomes:

$$w'(\varsigma) = A_w'(v_*)|x_*(\varsigma)|^\alpha (1-|\varsigma|)^\alpha \tag{23}$$

where $A_w'(v_*) = A_w(v_*)/w_n$.

The initial conditions (12) are also re-written in the normalized variables:

$$x_{*l}(t_0) = x_{*l0}, \ x_{*u}(t_0) = x_{*u0}, \quad w'(\varsigma, t_0) = w_0'(\varsigma) \quad -1 \leq \varsigma \leq 1 \tag{24}$$

Finally, we need to solve the system (17)-(20) on the interval (-1,1) under the initial conditions (24) with the stress contrast defined by (21), (22) and with the asymptotics of the opening given by (23).

## 4. Reduction to system of ordinary differential equations

The solution is found for the normalized quantities, in which we omit primes to simplify notation. The interval [-1, 1] of the normalized coordinate $\varsigma$ is represented with two sub-intervals [-1, 0) and [0,1]. The first of them is divided into $N_1$ elements of the same size $\Delta\varsigma_1 = 1/N_1$ and start points at $\varsigma_i = -1 + (i-1)\Delta\varsigma_1$ ($i = 1,..., N_1$); the second, into $N_2$ equal elements of the length $\Delta\varsigma_2 = 1/N_2$ and start points at точках $\varsigma_i = (i - N_1 - 1)\Delta\varsigma_2$ ($i = N_1 + 1,..., N$), where $N = N_1 + N_2$ is the total number of elements at the interval [-1, 1]. The lower tip ($\varsigma = -1$), the origin ($\varsigma = 0$) and the upper tip ($\varsigma = 1$) have the numbers, respectively, $i = 1$, $i = N_1 + 1$ and $i = N + 1$.

The discretized opening, particle velocity, differential pressure and leak-off term are prescribed as grid functions: $w_i = w(\varsigma_i)$, $v_i = v(\varsigma_i)$, $p_i = p_d(\varsigma_i)$, $q_{li} = q_l(\varsigma_i)$ ($i = 1,..., N+1$). The values of the opening are zero at the interval ends: $w_1 = 0$, $w_{N+1} = 0$. The values of the particle velocity at the ends of the interval equals to the propagation speed: $v_1 = v_{*l}$, $v_{N+1} = v_{*u}$.

For the source ($i = N_1 + 1$) given by delta-function, the volume balance accounting for the influx intensity $Q(t)$, is

$$\frac{dw_{N_1+1}}{dt} = \frac{3Q(t) - w_{N_1-1}v_{N_1-1} + 4w_{N_1}v_{N_1} - 4w_{N_1+2}v_{N_1+2} + w_{N_1+3}v_{N_1+3}}{2[\Delta\varsigma_1|x_{*l}| + \Delta\varsigma_2 x_{*u}]} \tag{25}$$

The particle velocity is discontinuous at the source. Thus, at the points $i = N_1$ and $i = N_1 + 2$, closest to the source, continuity equation is approximated to merely the first order of accuracy:

$$\frac{dw_{N_1}}{dt} = \frac{w_{N_1}(\varsigma_{N_1}|v_{*l}| - v_{N_1}) - w_{N_1-1}(\varsigma_{N_1-1}|v_{*l}| - v_{N_1-1})}{|x_{*l}|\Delta\varsigma_1} - \frac{v_{*l}}{x_{*l}}w_{N_1} - q_{lN_1} \tag{26}$$

$$\frac{dw_{N_1+2}}{dt} = \frac{w_{N_1+3}(\varsigma_{N_1+3}v_{*u} - v_{N_1+3}) - w_{N_1+2}(\varsigma_{N_1+2}v_{*u} - v_{N_1+2})}{x_{*u}\Delta\varsigma_2} - \frac{v_{*u}}{x_{*u}}w_{N_1+2} - q_{lN_1+2} \tag{27}$$

At the remaining division points ($i = 2, ..., N_1 - 1, N_1 + 3, ..., N$), central differences are available:

$$\frac{dw_i}{dt} = \frac{w_{i+1}(\varsigma_{i+1}|v_*| - v_{i+1}) - w_{i-1}(\varsigma_{i-1}|v_*| - v_{i-1})}{2|x_*|\Delta\varsigma(i)} - \frac{v_*}{x_*}w_i - q_{li} \tag{28}$$

where $\Delta\varsigma(i) = \begin{cases}\Delta\varsigma_1 \\ \Delta\varsigma_2\end{cases}$, $x_* = \begin{cases}x_{*l} \\ x_{*u}\end{cases}$, $v_* = \begin{cases}v_{*l} \\ v_{*u}\end{cases}$ $\begin{matrix}1 \leq i \leq N_1 \\ N_1 + 1 \leq i \leq N\end{matrix}$.

the system of ordinary differential equations (ODE) (25)-(28) contains $N - 1$ equations. It is complemented with two speed equations (19), which being discretized become:

$$\frac{dx_{*l}}{dt} = -\left(\frac{1}{t_s}\right)^{1/n}\left(\frac{1}{A_\mu}\frac{w_2}{|x_{*l}\Delta\varsigma_1|^\alpha}\right)^{\frac{1}{1-\alpha}}, \qquad \frac{dx_{*u}}{dt} = \left(\frac{1}{t_s}\right)^{1/n}\left(\frac{1}{A_\mu}\frac{w_N}{|x_{*u}\Delta\varsigma_2|^\alpha}\right)^{\frac{1}{1-\alpha}} \qquad (29)$$

Therefore, we have the dynamic system of $N+1$ ODE (25)-(29) in $N+1$ unknowns; the first $N-1$ equations contain the temporal derivatives of the unknown openings $w_i$ at the internal nodes $i = 2, \ldots, N_1 - 1, N_1 + 3, \ldots, N$ of the mesh, while the last two contain the derivatives of unknown coordinates $x_{*l}$ and $x_{*u}$ of fracture tips, which are also present in the right hand sides of equations (25)-(28). The values $v_{*l}$ and $v_{*u}$ entering (26)-(28) are defined by the right hand sides of the speed equations (29).

The nodal values of the particle velocity, needed to evaluate right hand sides of ODE (25)-(28), are found from discretized Poiseuille type equation (18). The nodal values of the velocity are evaluated by using modified formula of central differences [27], which accounts for the asymptotics of the differential pressure, which behaves as $O(r^{\alpha-2})$, where $r$ is the distance form a tip, $0 < \alpha < 1$. As a result, for nodal values of the velocity we obtain:

$$v_1 = v_{*l}, \quad v_2 = 0.5(v_{*l} + v_3),$$

$$v_i = -sign(p_{i+1} - p_{i-1})\left(w_i^{n+1}\left|\frac{p_{i+1}-p_{i-1}}{x_*\Delta f}\right|\right)^{1/n} \qquad i = 3, \ldots, N_1, N_1 + 2, \ldots, N-1 \qquad (30)$$

$$v_N = 0.5(v_{*u} + v_{N-1}), \quad v_{N+1} = v_{*u}$$

where $\Delta f = \frac{(1-|\varsigma_i|)^{\alpha(n+1)}}{\alpha(n+1)-1}\left[\frac{1}{(1-|\varsigma_{i+1}|)^{\alpha(n+1)-1}} - \frac{1}{(1-|\varsigma_{i-1}|)^{\alpha(n+1)-1}}\right]$. Note that the particle velocity at the source ($i = N_1 + 1$) is undefined and it is not used in ODE (25)-(29).

The nodal values of the differential pressure entering (30) are found from the discretized elasticity equation (20) with approximation of the density near a node $j$ as $w(\varsigma) = \frac{w_j}{(1-|\varsigma_j|)^{2/3}}(1-|\varsigma|)^{2/3}$, which accounts for the asymptotics (23) for a Newtonian fluid ($n = 1$, $\alpha = 2/3$). This serves to accurately approximate the opening both near and far from the front for any thinning fluid ($0 < n < 1$). Under this approximation, hypersingular integral entering (20) is evaluated analytically over an arbitrary interval $(a, b)$. Specifically, for $a, b < 0$, we have $\int_a^b \frac{(1-|\tau|)^{2/3}}{(\tau-\omega)^2}d\tau = \Phi(\sqrt[3]{-a}, \sqrt[3]{-\omega}) - \Phi(\sqrt[3]{-b}, \sqrt[3]{-\omega})$, while for $a, b > 0$, we have $\int_a^b \frac{(1-|\tau|)^{2/3}}{(\tau-\omega)^2}d\tau = \Phi(\sqrt[3]{b}, \sqrt[3]{\omega}) - \Phi(\sqrt[3]{a}, \sqrt[3]{\omega})$, where $\Phi(\xi, z)) = \frac{\xi^2}{z^3-\xi^3} + \frac{2}{3}\frac{1}{z}\left[\frac{1}{2}\ln\left|\frac{(\xi/z-1)^3}{(\xi/z)^3-1}\right| + \sqrt{3}arctg\frac{1+2\xi/z}{3}\right]$. For the node $j = 2$, integration is performed over the interval $(-1, -1 + 1.5\Delta\varsigma_1)$; for the source ($j = N_1 + 1$), over two intervals $(-0.5\Delta\varsigma_1, 0)$ and $(0, 0.5\Delta\varsigma_2)$; for remaining nodes, over the interval $(\varsigma_j - \Delta\varsigma/2, \varsigma_j + \Delta\varsigma/2)$ for the $j$-th node. Finally, for the nodal values of the differential pressure, we have:

$$p_i = \frac{1}{4\pi}\sum_{j=2}^N A_{ij}(w_j + w_{SCj}) \qquad i = 2, \ldots N \qquad (31)$$

In (31), the influence coefficients $A_{ij}$ are obtained as explained by integration over the interval corresponding to the number $j$. The values $w_{SCj} = w_{SC}(\varsigma_j)$ present the nodal values of the opening $w_{SC}(\varsigma)$, generated by piece-wise constant stress contrast. This opening is defined by equation following form the classical solution [29] of the elasticity problem for a straight crack under arbitrary traction. For the piece-wise constant stress contrast (21), it is:

$$w_{SC}(\varsigma) = \left(\frac{4}{\pi}\right)\{\Delta\sigma^l[g(\varsigma,\varsigma^l) - g(\varsigma,-1)] + \sum_{s=l+1}^{s=u-1}\Delta\sigma^s[g(\varsigma,\varsigma^s) - g(\varsigma,\varsigma^{s-1})] +$$

$$\Delta\sigma^u[g(\varsigma,1) - g(\varsigma,\varsigma^{u-1})]\} \tag{32}$$

where $g(\varsigma,\tau) = (\xi_1 - x_1)\text{arch}\left|\frac{x_m^2 - x_1\xi_1}{x_m(\xi_1 - x_1)}\right| - \sqrt{x_m^2 - x_1^2}\arccos\frac{\xi_1}{x_m}$, $x_m = 0.5(x_{*u} - x_{*l})$, $\xi_1 = \xi_1(\tau) = |x_*(\tau)|\tau - x_0$, $x_1 = x_1(\varsigma) = |x_*(\varsigma)|\varsigma - x_0$, $x_0 = 0.5(x_{*u} + x_{*l})$, $l$ and $u$ are the numbers of layers containing the lower and upper tips of the fracture, respectively; $\varsigma^s$ is the normalized coordinate of the $s$-th contact defined by (22). Summation under the sum symbol in (32) is performed only when $u > l + 1$. In the case, when $u = l$, the both tips are in the pay layer, there is no stress contrast ($\Delta\sigma^l = \Delta\sigma^u = 0$) and consequently $w_{SC}(\varsigma) = 0$. In this case, the solution is that of the classical KGD problem, for which one can use the highly accurate bench mark solution (e. g. [25]). Note that equation (32) yields that at contacts of layers, the function $w_{SC}(\varsigma)$ is continuous while its derivative has log-type singularity.

Summarizing, the right hand sides of the system (25)-(29) are evaluated as follows. Firstly, for given at a considered moment nodal openings $w_i$ and coordinates $x_{*l}$ and $x_{*u}$ of the tips, we use equation (31) to find the nodal values of the differential pressure $p_i$. The latter are substituted into (30) to evaluate the nodal values of the particle velocity, which in their turn are substituted into continuity equations (25)-(28). Therefore, at each time instant, the right hand sides of equations (25)-(29) are defined. Thus integration of the ODE (25)-(29) under the discretized initial conditions (24) may be performed by means of well-developed methods (see, e. g. [30]).

## 5. Numerical results and their discussion

The described computational scheme is implemented for an arbitrary system of layers with arbitrary stress contrast and any numbers $N_1$ and $N_2$ points of the fracture division. The power-law fluid may have an arbitrary behaviour index $n$. The initial conditions can be arbitrary, as well. Solution of the Cauchy problem for the system of ODE is performed by implicit backward differentiation method [31].

Below for certainty we present results for the case of a Newtonian fluid ($n = 1$), constant influx at the source ($Q(t) = const$) and impermeable rock ($q_l = 0$). The initial conditions are those of the classical KGD problem: zero fracture length and zero opening on it at the initial time ($x_{*l}(0) = 0$, $x_{*u}(0) = 0$, $w_0(0) = 0$). Therefore, until the fracture reaches a boundary of the pay layer, the problem is self-similar with known self-similar solution (see, e. g. [11,16,22,25]). By special tests with various number of division points it was established that the numbers of points $N_1 = N_2 = 51$ guarantee the accuracy 1.3%, at least, for all the values discussed below.

After normalizing, the external characteristics of the problem under prescribed geometry of layers are merely the influx $Q$ and stress contrast $\sigma(x)$. From physical considerations it is clear that passing through a contact occurs differently depending on whether the fracture penetrates layer with higher ($\sigma^+ > \sigma^-$) or lower ($\sigma^+ < \sigma^-$) stresses. In the former case, we call the contact positive, in the second negative. Clearly, peculiarities of the transition through a positive or negative contact are defined by the difference $\sigma^+ - \sigma^-$ in the neighboring layers. Therefore, the study of the peculiarities may be performed for the simplest symmetric scheme (Fig. 2) when the pay layer with the stress $\sigma^- = \sigma_0$ is located between half-planes with the same compressive stress $\sigma^+$. Then due to the symmetry of the problem, the number of unknowns reduces two-fold, while the transition through contacts under prescribed thickness $H$ of the pay layer is entirely characterized by the stress contrast $\Delta\sigma = \sigma^+ - \sigma_0$.

Calculations for this scheme are performed in the normalized variables, while their results are discussed in non-normalized quantities for the particular case considered in the paper [24]. Specifically, the following parameters of the model are used: $H$ = 50 m, $E'$ = 2.5 $10^4$ MPa, $\mu'$ = 1.2 Pa s, $Q_0$ = 5 $10^{-4}$ m$^2$/s. For stress contrast, the authors of [24] assigned three positive values: $\Delta\sigma$ = 0.5, 1.5 and 3.5 MPa. To study the transition in detail, we additionally consider (i) the case of extremely high positive barrier ($\Delta\sigma$ = 50 MPa) and (ii) previously non-studied cases of negative barriers ($\Delta\sigma$ = - 0.5, - 1.5 and -3.5 MPa).

For the positive stress contrasts $\Delta\sigma$, studied in [24], the fracture length $2x_* = 2x_{*u}$ and the opening profiles agree with those of the cited paper to the accuracy of graphs given in it. For the positive stress contrast $\Delta\sigma$ =50 MPa, notably higher than the maximal contrast $\Delta\sigma$ = 3.5 MPa studied in [24], the results are presented in Fig. 3. It shows the changes in time of the fracture half-length $x_*$ (Fig. 3a), the average pressure $p_{av}$ (Fig. 3b), the average opening $w_{av}$ (Fig. 3c) and the propagation speed $v_*$ (Fig. 3d). It can be seen that the fracture practically does not propagate until the average differential pressure in it is an order less than the stress contrast $\Delta\sigma$. In the considered case, the delay time is about 3000s. Fig. 3c implies that during this time, the average opening increases proportionally to the fluid volume entering the fracture. The propagation speed $v_*$ (Fig. 3d), when reaching high barrier, drops two orders (from 0.1 m/s to 0.001 m/s) and oscillates about the small value until the pressure $p_{av}$ becomes of order of stress contrast $\Delta\sigma$.

The oscillations of the propagation speed $v_*$ arise because fluid particles, when reaching high barrier, stop and even change the direction of motion near fracture tips. The particle velocity profiles, shown in Fig. 4, illustrate this complicated motion for the stress contrast $\Delta\sigma$ = 5 MPa. It can be seen that after the velocity drops practically to zero at the moment $t$ = 201.4 s, there occurs back flow near the fracture tip in the time interval from 201.7 s to 210 s. As show detailed calculations, fluctuations of the particle velocity near the barrier continue after that; they damp merely when $t > 10^4$ s.

It is of interest to reveal to what extent the stress jump influences the accuracy of the results. To this end, the calculations for the stress contrast $\Delta\sigma$ =50 MPa are repeated with change of the step-function to continuous distribution of $\Delta\sigma$ from zero to 50 MPa in a narrow strip of 2 m width located symmetrically about the boundary $x = H/2$ = 25 m. In these calculations, we took a fine mesh near the barrier. Typical results for step-wise change of stress contrast are shown in Fig. 5 at the time $t$ = 12000 с. They are indistinguishable from those obtained in calculations for the contrast prescribed by step-function. The agreement of the results justifies assigning stress contrast by step-function.

Fig. 6 presents results for boundaries with *negative* stress contrasts ($\Delta\sigma$ = -0.5, -1.5 and -3.5 MPa) in comparison with those for positive stress contrasts of the same magnitude. Clearly, in difference with positive contrast, the transition through negative contrast leads to drastic acceleration of the fracture and to drop of average pressure in it. Meanwhile, similar to positive contrast, with time growth the propagation speed asymptotically tends to the value corresponding to the classical KGD model ($\Delta\sigma$ = 0), while the average differential pressure approaches the assigned stress contrast $\Delta\sigma$. Note that the effect of faster propagation in a layer with decreased confining pressure has been observed in 3D modeling with the implicit level set algorithm [28].

From physical considerations and numerical results for both positive and negative contrasts it follows that the internal characteristic of the transition through a contact serves the difference between the fluid and rock pressures in the fracture at the moment when a tip reaches the contact. Denote the average difference $P$. For the pay layer, $P$ is the average net-pressure at this moment. Recalling that the external characteristic is the difference $\Delta\sigma = \sigma^+ - \sigma^-$, we infer that the transition is controlled by the dimensionless parameter

$$R = \frac{\Delta\sigma}{P} \tag{33}$$

It can be promptly evaluated for the simplest symmetric scheme of Fig. 2. In this case, the moment $t_H$ of reaching the contact is defined by the condition $x_*(t_H) = H/2$, while all the quantities are known for the self-similar solution studied in detail in [19,24,30]. By employing this solution, we obtain the analytical expression for the average differential pressure $P$ at the moment of reaching the contact:

$$P = \frac{\sqrt[4]{Q\mu' E'^3}}{k_0 \sqrt{H}} \tag{34}$$

with $k_0 = 1/(P_0\sqrt{2\xi_{*n}})$, and for a Newtonian fluid ($n = 1$) $P_0 = 0.3832$, $\xi_{*n} = 0.6113$, thus $k_0 = 2.360$. Substitution of (34) into (33) yields equation for the contrast parameter $R$ in the considered particular case:

$$R = k_0 \frac{\Delta\sigma\sqrt{H}}{\sqrt[4]{Q\mu' E'^3}} \tag{35}$$

Note that for positive (negative) stress contrast, the contrast parameter $R$ is positive (negative). For positive contrast, by taking (35) squared, we can write it as $R^2 = k_0^2 \tilde{H}$, where $\tilde{H}$ is the parameter introduced in the paper [24] for the scheme of Fig. 2:

$$\tilde{H} = \frac{\Delta\sigma^2 H}{\sqrt{Q\mu' E'^3}}$$

Surely, it is easy to extend the parameter $\tilde{H}$ to negative contrast by including the factor sign $\Delta\sigma$ into its definition. However, the parameter $\tilde{H}$, being practically equivalent to the parameter $R$ in the particular case discussed, is inapplicable in the general case.

In conclusion we give an example illustrating application of the method developed to a non-symmetric system of layers with different stress contrasts. Fig. 7 presents profiles of the opening at various time instances, when below and above the pay layer of the thickness $H = 50$ m there are layers of various thicknesses and different stress contrasts. The elasticity modules, fluid viscosity and its influx are those assumed above for the scheme of Fig. 2. It can be seen that high positive barrier ($\Delta\sigma = 7.5$ MPa) practically arrests the fracture, which after overcoming the barrier with less stress contrast ($\Delta\sigma = 5$ MPa) rapidly propagates in the layer where the contrast drops to zero.

**Acknowledgement.** The authors appreciate the support of the Russian Scientific Fund (Grant # 15-11-00017).

## References

1. Adachi J., Siebrits E., Pierce A., Desroches J. Computer simulation of hydraulic fractures // *Int. J. Rock Mech. Mining Sci.* 2007. V. 44. N 5. P. 739─757.

2. *Effective and Sustainable Hydraulic Fracturing*. Eds A. P. Bunger et al. Proc. Int. Conf. HF─2013. Croatia: InTech. 1000 p. www.intechopen.com.

3. Mack M.G., Warpinski N.R. Mechanics of hydraulic fracturing. // *Reservoir simulation*, M. J. Economides, K.G. Nolte (eds), 3-rd edn. John Willey & Sons, 2000, Chapter 6.

4. Khristianovich S.A., Zheltov V.P. Formation of vertical fractures by means of highly viscous fluid // *Proc. 4-th World Petroleum Congr.* Rome, 1955. P. 579—586.

5. Желтов Ю.П., Христианович С.А. О гидравлическом разрыве нефтеносного пласта // *Изв. АН СССР, ОТН*. 1955. № 5. С. 3—41 (in Russian).

6. Geertsma J., de Klerk F. A rapid method of predicting width and extent of hydraulically induced fractures // *J. Pet. Tech.* 1969. V. 21. P. 1571—1581.

7. Perkins T.K., Kern L.R. Widths of hydraulic fractures // *J. Pet. Tech*. 1961. V. 13. N 9. P. 937—949.

8. Howard G.C., Fast C.R. *Hydraulic Fracturing*. SPE of AIME. Dallas, 1970. 203 p.

9. Nordgren R.P. Propagation of a vertical hydraulic fracture // *Soc. Pet. Eng. Journal*. 1972. V. 12. P. 306—314.

10. Settari A, Cleary M. Development and testing of a pseudo-three-dimensional model of hydraulic fracture geometry (P3DH) // *Proc. 6-th SPE symposium on reservoir simulation of the society of petroleum engineers*. 1982. SPE 10505. P. 185–214.

11. Spence D.A., Sharp P.W. Self-similar solutions for elastohydrodynamic cavity flow // *Proc. Roy Soc. London. Ser. A*. 1985. V. 400. N 819. P. 289—313.

12. Kemp L.F. Study of Nordgren's equation of hydraulic fracturing // *SPE Production Eng.* 1990. V. 5. P. 311—314.

13. Desroches J., Detournay E., Lenoach B., Papanastasiou P., Pearson J.R.A., Thiercelin M., Cheng A. The crack tip region in hydraulic fracturing // *Proc. Roy Soc. London. Ser. A*. 1994. V. 447. N 1929. P. 39—48.

14. Lenoach B. The crack tip solution for hydraulic fracturing in a permeable solid // *J. Mech. Phys. Solids*. 1995. V. 43. P. 1025—1043.

15. Garagash D.I., Detournay E. The tip region of a fluid-driven fracture propagating in an elastic medium // *Trans. ASME J. Appl. Mech.* 2000. V. 67. N 1. P. 183—192.

16. Adachi J., Detournay E. Self-similar solution of plane-strain fracture driven by a power-law fluid // *Int. J. Numer. Anal. Meth. Geomech*. 2002. V. 26. P. 579-604.

17. Savitski A., Detournay E. Propagation of a penny-shaped fluid driven fracture in an impermeable rock: asymptotic solutions // *Int. J. Solids Struct*. 2002. V. 39. N 26. P. 6311—6337.

18. Jamamoto K., Shimamoto T., Sukemura S. Multi fracture propagation model for a three-dimensional hydraulic fracture simulator // *Int. J. Geomech. ASCE*. 2004. V. 4. N 1. P. 46—57.

19. Peirce A., Detournay E. An implicit level set method for modeling hydraulically driven fractures // *Comput. Methods Appl. Mech. Engng*. 2008. V. 197. P. 2858—2885.

20. Rahman M.M., Rahman M.K. A review of hydraulic fracture models and development of an improved pseudo-3D model for stimulating tight oil/gas sand // *Energy Sources, Part A: Recovery, Utilization, and Environmental Effects*. 2010. V. 32. N 15. P. 1416-1436.

21. Linkov A.M. Speed equation and its application for solving ill-posed problem of hydraulic fracturing // *Doklady Physics*. 2011. V. 56. N 8. P. 436-438.

22. Linkov A.M. On efficient simulation of hydraulic fracturing in terms of particle velocity // *Int. J. Engineering Sci.* 2012. V. 52. P. 77-88.

23. Gordeliy E., Peirce A. Implicit level set schemes for modeling hydraulic fractures using the XFEM // *Computer Meth. Appl. Mech. Eng*. 2013. V. 266. P. 125–143.

24. Dontsov E.V., Peirce A.P. An enhanced pseudo-3D model for hydraulic fracturing accounting for viscous height growth, non-local elasticity, and lateral toughness // *Engineering Fracture Mechanics.* 2015. V. 142. P. 116–139.

25. Wrobel M., Mishuris G. Hydraulic fracture revisited: Particle velocity based simulation // *Int. J. Eng. Sci.* 2015. V. 94. P. 23-58.

26. Linkov A.M. Particle velocity, speed equation and universal asymptotics for efficient modelling of hydraulic fractures // *J. Appl. Math. Mech.* 2015. V. 79. P. 54–63.

27. Linkov A.M. Numerical solution of hydraulic fracture problem in modified formulation under arbitrary initial conditions // *J. Mining Sci.* 2016. V. 52. N 2. P. 265-273.

28. Peirce A. Implicit level set algorithms for modeling hydraulic fracture propagation // *Phil. Trams. R. Soc.* A 374: 20150423, 2016.

29. Muskhelishvili N.I. *Some Basic Problems of the Mathematical Theory of Elasticity.* Groningen, Noordhoff. 1975. 746 p.

30. Epperson J.F. An *Introduction to Numerical Methods and Analysis* . J. Willey & Sons, Inc. 2007. 572 p.

31. Brayton R.K., Gustavson F.G., Hachtel G.D. A new efficient algorithm for solving differential-algebraic systems using implicit backward differentiation formulas // *Proc. IEEE.* 1972. V. 60. N. 1. P. 98-108.

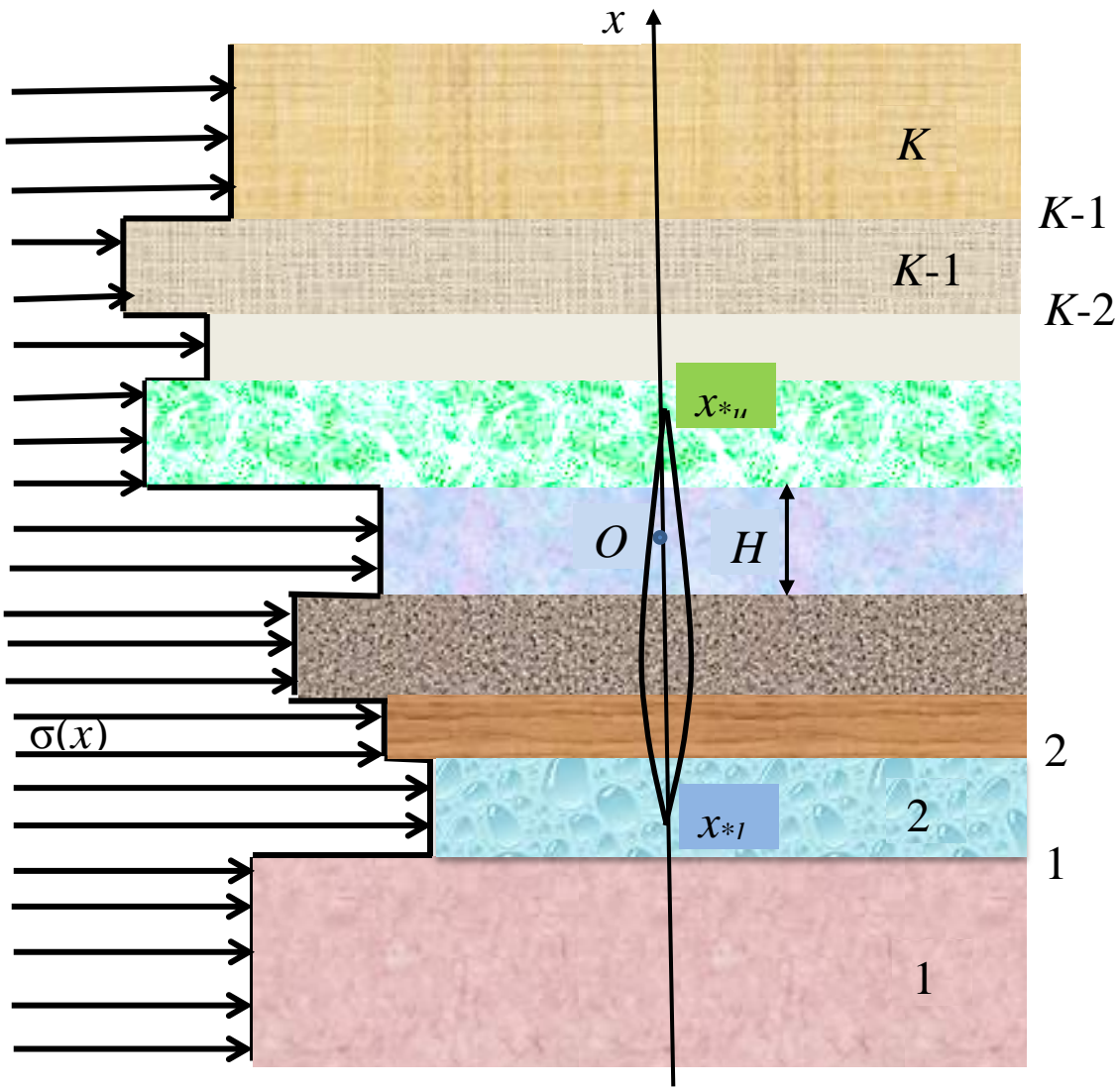

Fig. 1. The scheme of a hydraulic fracture in layered rock with various stresses in layers

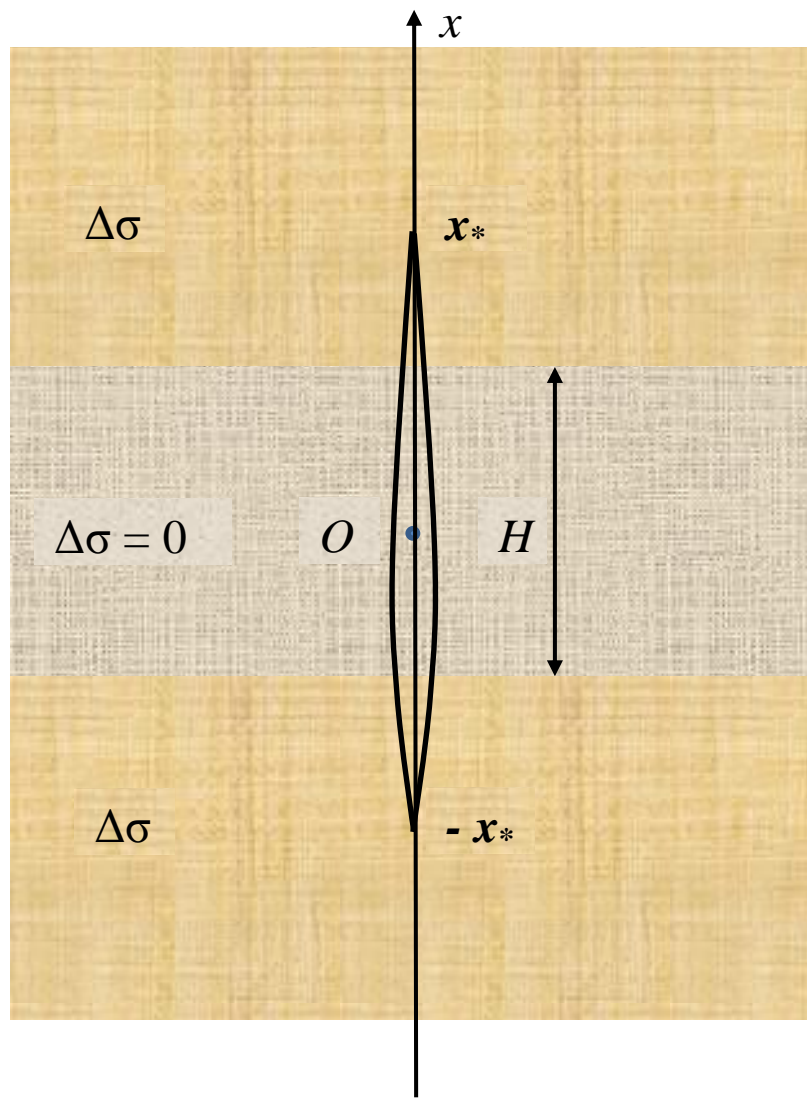

Fig. 2. The symmetric scheme of a pay layer between half-planes with the same stress contrast

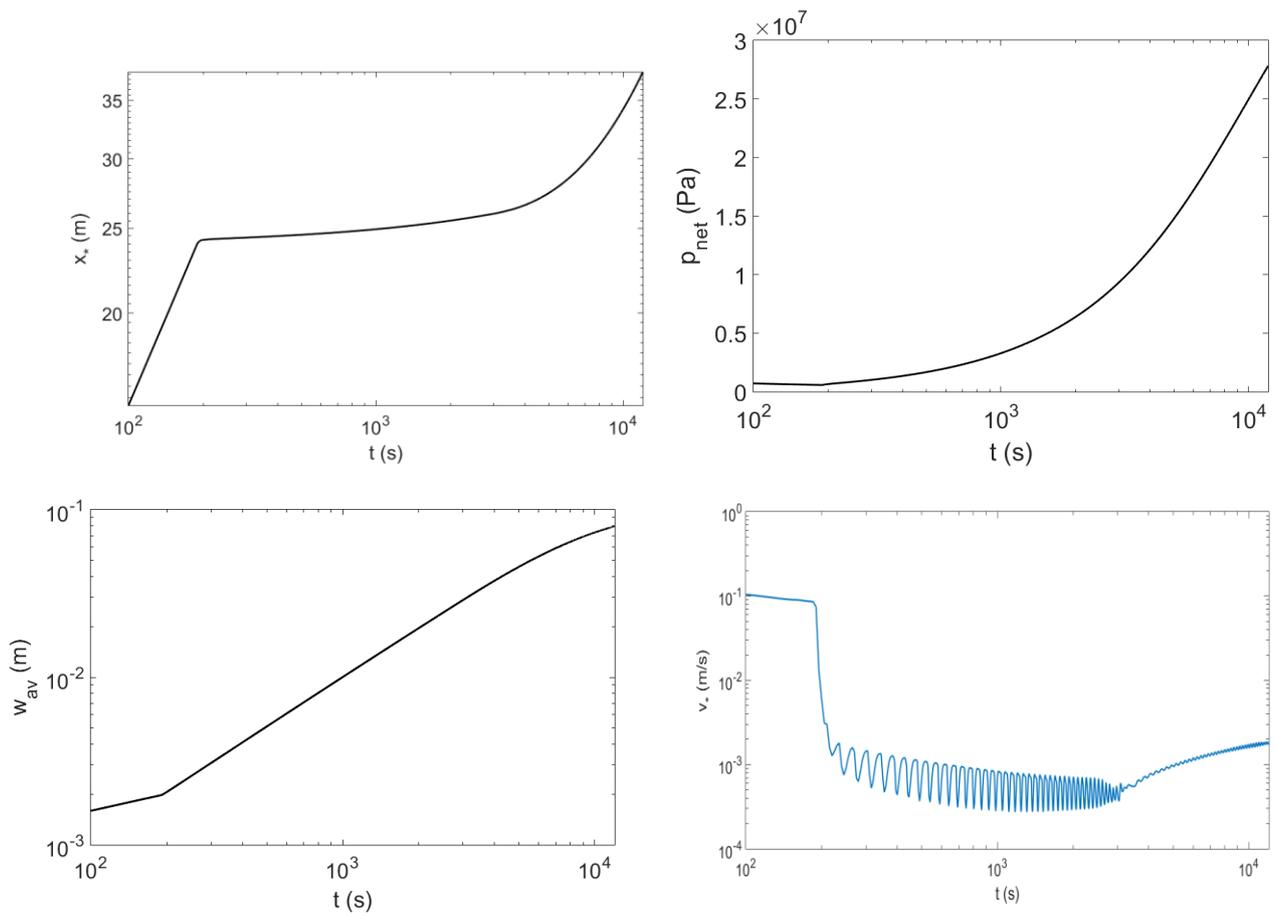

Fig. 3. The change in time of a) half-length, b) average net-pressure, c) average opening, d) propagation speed when the fracture passes through the barrier with stress contrast $\Delta\sigma = 50$ MPa

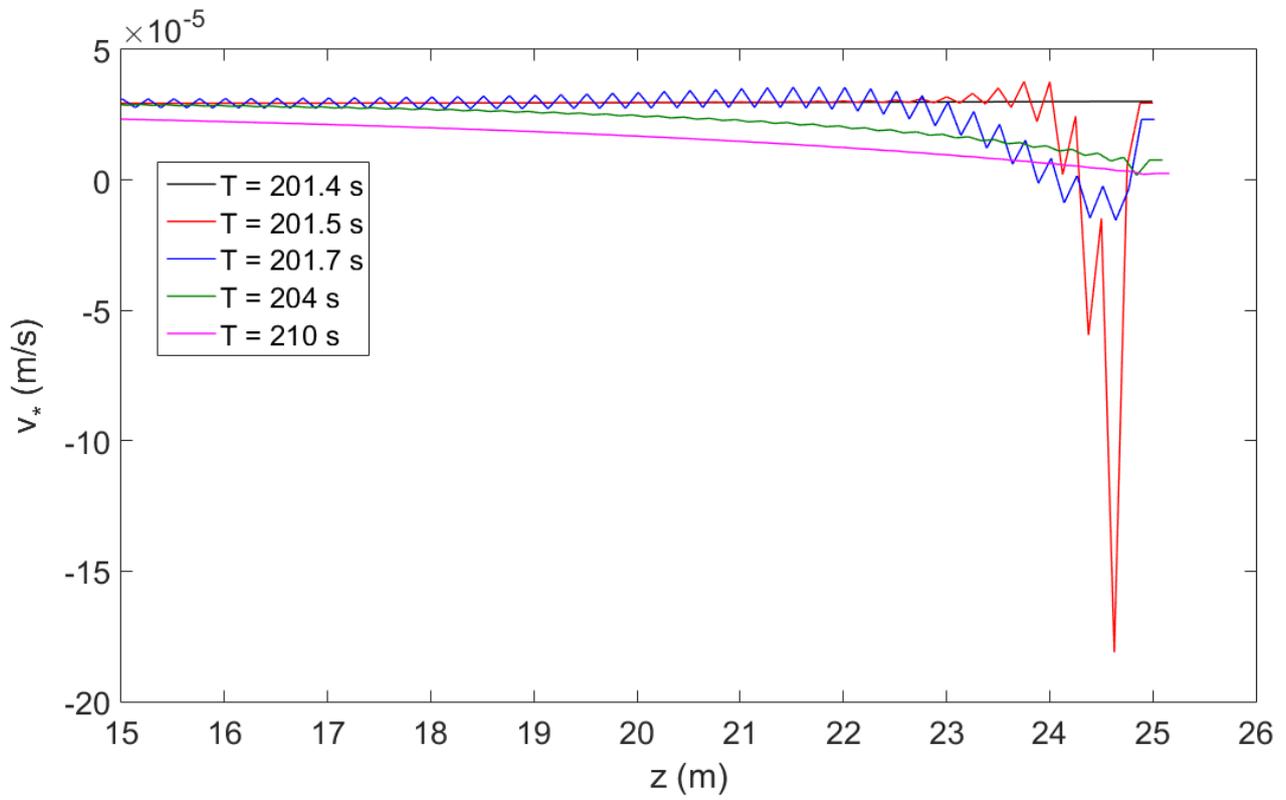

Fig. 4. Profiles of the fluid velocity in the fracture penetrating the barrier with stress contrast Δσ = 5 MPa

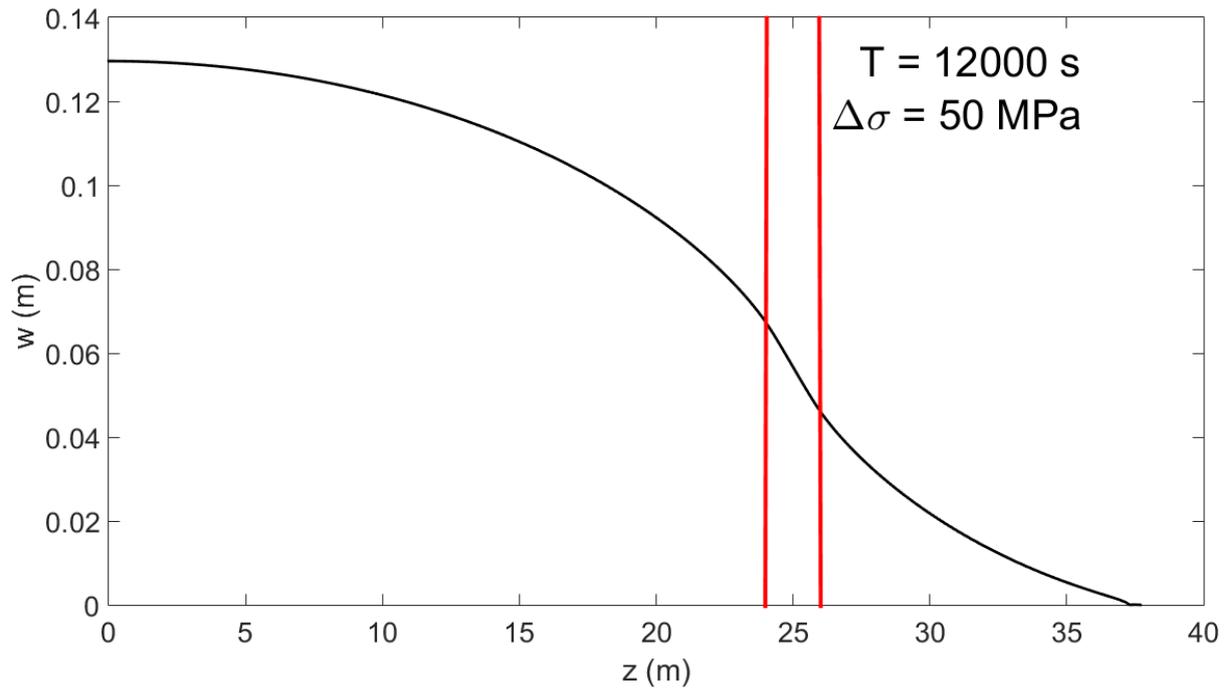

Fig. 5. Profile of the opening for a barrier represented by linear change of stress contrast within a strip of 2m width. It is indistinguishable from that for the contrast prescribed by step-function

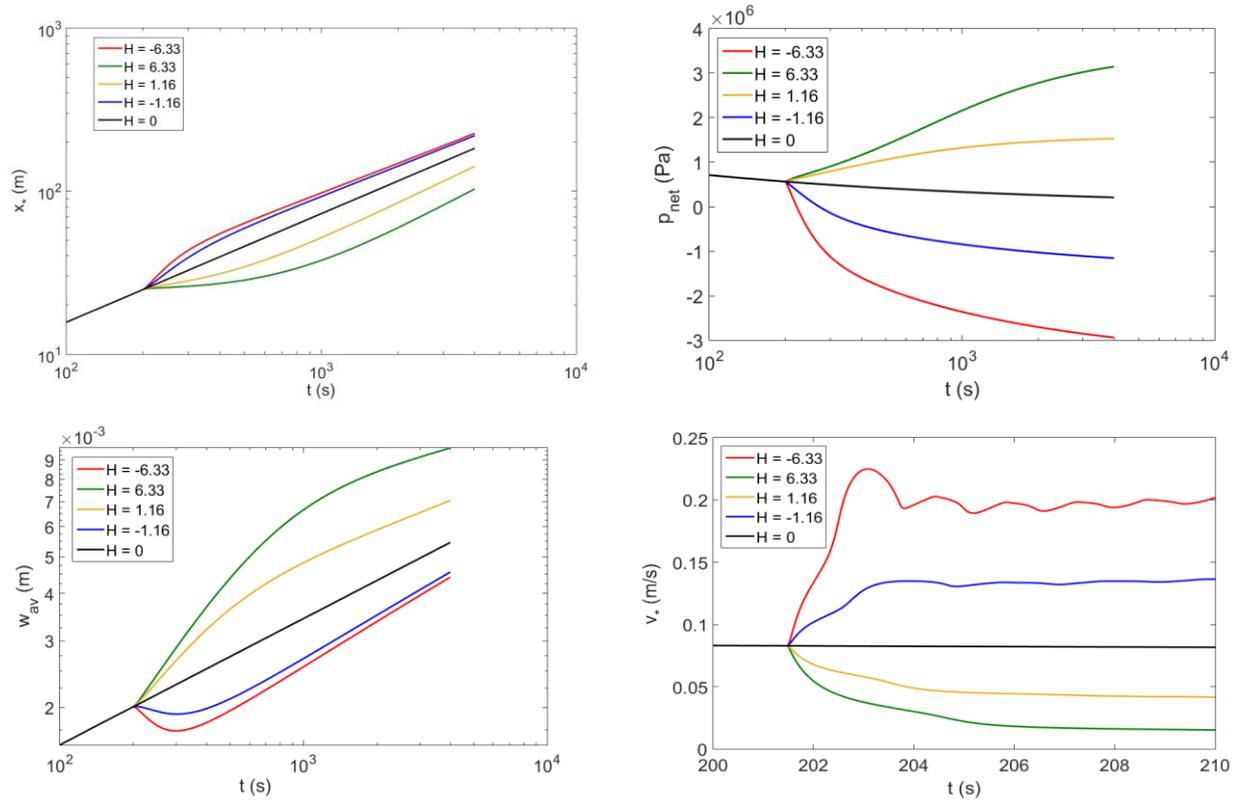

Fig. 6. The change in time of a) fracture half-length, b) average net-pressure, c) average opening and d) particle velocity for barriers with positive ($\tilde{H} > 0$) and negative ($\tilde{H} < 0$) stress contrasts

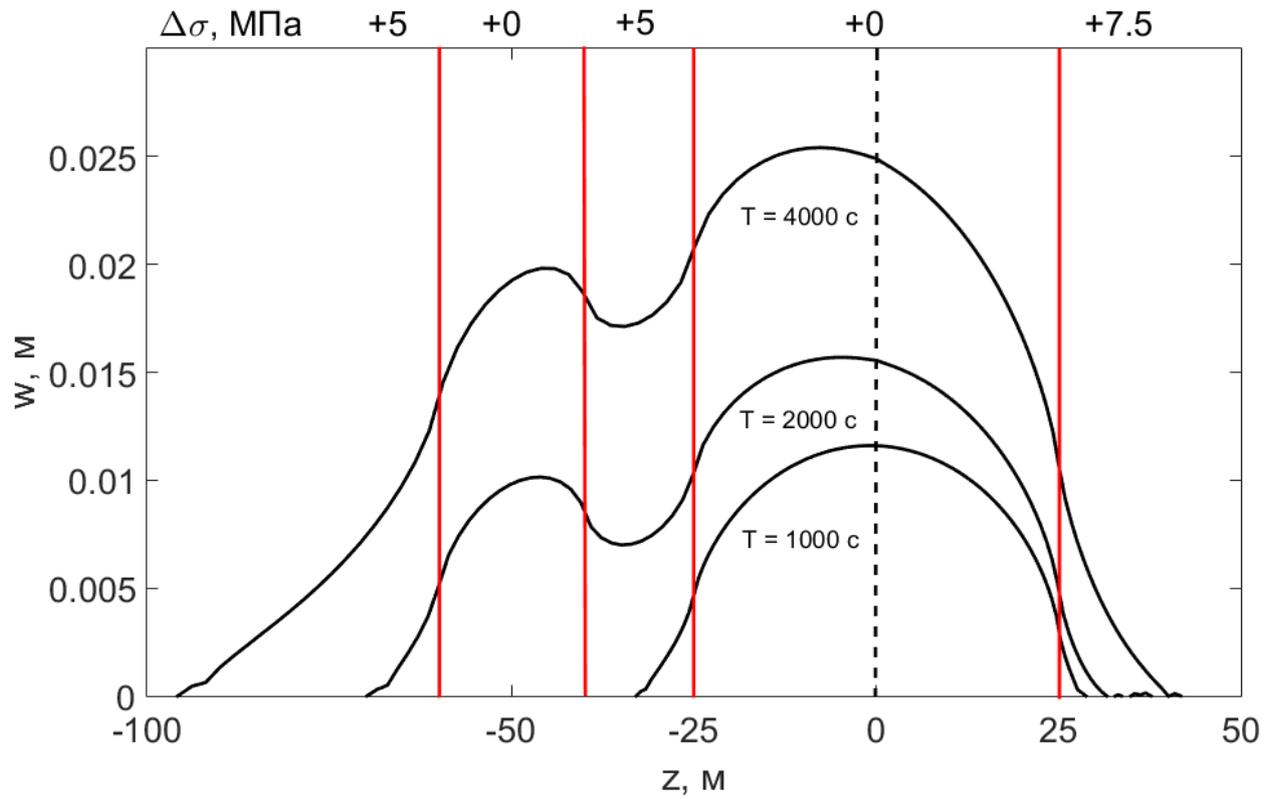

Fig. 7. Profiles of the opening at various time instances for a non-symmetric system of layers with various stress contrasts (values of Δσ are shown on top of the figure)